\documentclass[twocolumn,trackchanges]{aastex631}

\usepackage{mathptmx}
\usepackage[T1]{fontenc}
\usepackage{ae,aecompl}
\usepackage{graphicx}	
\usepackage{amsmath}	
\usepackage{amssymb}	

\begin{document}

\graphicspath{{./}{figures/}}

\title{Ejected Particles after Impact Splash on Mars: Electrification}

\author{T. Becker}
\affiliation{University of Duisburg-Essen, Faculty of Physics, Lotharstr. 1, 47057 Duisburg, Germany}

\author{F.C. Onyeagusi}
\affiliation{University of Duisburg-Essen, Faculty of Physics, Lotharstr. 1, 47057 Duisburg, Germany}

\author{J. Teiser}
\affiliation{University of Duisburg-Essen, Faculty of Physics, Lotharstr. 1, 47057 Duisburg, Germany}

\author{T. Jardiel}
\affiliation{Instituto de Cer\'{a}mica y Vidrio, CSIC, C/Kelsen 5, Campus Cantoblanco 28049 Madrid, Spain}

\author{M. Peiteado}
\affiliation{Instituto de Cer\'{a}mica y Vidrio, CSIC, C/Kelsen 5, Campus Cantoblanco 28049 Madrid, Spain}

\author{O. Mu{\~n}oz}
\affiliation{Instituto de Astrof\'{i}sica de Andaluc\'{i}a, CSIC Glorieta de la Astronom\'{i}a s/n 18008 Granada, Spain}

\author{J. Martikainen}
\affiliation{Instituto de Astrof\'{i}sica de Andaluc\'{i}a, CSIC Glorieta de la Astronom\'{i}a s/n 18008 Granada, Spain}

\author{J.C. Gomez Martin}
\affiliation{Instituto de Astrof\'{i}sica de Andaluc\'{i}a, CSIC Glorieta de la Astronom\'{i}a s/n 18008 Granada, Spain}

\author{J. Merrison}
\affiliation{Institute of Physics and Astronomy, Aarhus Universitet, 8000 Aarhus, Denmark}

\author{G. Wurm}
\affiliation{University of Duisburg-Essen, Faculty of Physics, Lotharstr. 1, 47057 Duisburg, Germany}

\begin{abstract}
Within the RoadMap project we investigated the microphysical aspects of particle collisions during saltation on the Martian surface in laboratory experiments. Following the size distribution of ejected particles, their aerodynamic properties and aggregation status upon ejection, we now focus on the electrification and charge distribution of ejected particles. We analyzed rebound and ejection trajectories of grains in a vacuum setup with a strong electric field of 100 kV/m and deduced particle charges from their acceleration. The ejected particles have sizes of about 10 to 100 microns. They carry charges up to $10^5 \, \rm e$ or charge densities up to $> 10^7 \rm e/mm^2$. Within the given size range, we find a small bias towards positive charges.
\end{abstract}

\section{Introduction}
Mars, as nearest terrestrial neighbor
has been the center of research for many years now. 
While manned missions are just coming into reach, many unmanned missions have provided data on properties accessible to remote sensing by orbiters or locally by rovers. These include soil composition \citep{Toulmin1977, Banin1979, Clark1981, Clark1982, McCord1982, Banin1986, Arvidson1989, Guinness1989}, past and recent particle movement on the surface \citep{Metzger1999, Greeley2006, Silvestro2010, Bridges2012b, Bridges2012, Reiss2016a, Chojnacki2011, Chojnacki2019, Waza2023}, and atmospheric processes \citep{Pollack1995, Tomasko1999, Wolff2003, Clancy2003, Lemmon2004, Wolff2006, Vandaele2019, Neary2020, Holmes2021, Daversa2022}. 

There are still many phenomena though, we have yet to fully understand. It is, e.g., well known that there are frequent local and global dust storms occurring on Mars \citep{Haberle1982, Cantor1999, Gurwell2005, Wang2015, Kass2019, Lemmon2019, Elsaid2020, Holmes2021}. However, there is still an ongoing debate on how that fine dust gets into the atmosphere in the first place. Direct dust entrainment \citep{Bagnold1954, Chepil1963, Shao1993, White1997, Loosmore2000, Kjelgaard2004, Roney2004, Macpherson2008}, vortex supported lifting by dust devils \citep{Thomas1985, Balme2006, Neakrase2010, Stanzel2008, Neakrase2016, Bila2020, Baker2021, Lorenz2021}, saltation release \citep{Greeley1976, Greeley1980, Iversen1976, Iversen1982, Merrison2012, kok2014, Musiolik2018, Swann2020, Kruss2020, Becker2022, Becker2023, Waza2023}, and thermal creep support \citep{deBeule2014, deBeule2015, Kuepper2016, Schmidt2017, Bila2024} may be the most prominent ones. All of them might apply to dust lifting on Mars. To what extent is unclear, though.

To shed light on the soil atmosphere interaction, our goal during the RoadMap project were laboratory experiments on dust lifting. One specific series of experiments considered individual impacts during saltation. So we focused on the microscopical scale, with the aim being a characterization of different aspects of ejecta during single impacts of grains into soil. In a first work, we quantified the size distribution of ejecta in the 1-5 micron range \citep{Becker2022}. In a second work, we analyzed the aerodynamic properties of ejected aggregates \citep{Becker2023}. Now, in this third work, we take a closer look on the electrification of ejected particles. 

When two non-conducting surfaces get into contact with each other or are seperated, they can exchange charge \citep{Gilbert1991, Lacks2019}. This tribocharging is a commonly known phenomenon but the mechanisms behind it seem elusive \citep{Lacks2019}. Nevertheless, tribocharging has been studied throughout many scientific communities \citep{Knoblauch1902, Shaw1926, Henniker1962, Lowell1988, Schein1999, Lacks2011,  Steinpilz2020, Jungmann&Wurm2021}. 

There are some general trends. Material dependent polarities occur frequently, known as triboelectric series \citep{Shaw1926, Shinbrot2008, Apodaca2010, Pham2011, Jungmann2018}. Details depend on many parameters though and can be related to surface properties, humidity, or contact history \citep{Lowell1986, Forward2009a, Ireland2010, Waitukaitis2014, Xie2016, Lee2018, Grosjean2023}.
Grain sizes regularly play a role, i.e. in collisions of same material particles, grains with smaller size tend to charge negatively compared to the large grains gathering positive charge \citep{Lacks2008, Duff&Lacks2008, Waitukaitis2014, Toth2017, Forward2009a}

Applied to saltation, this means that for every contact of a saltating grain with the soil, or any contact within the soil, the respective particles will gain a (small) charge. Depending on how large this charge is and if there is a pattern on how it is distributed among differently sized particles, charging could have a non-negligible effect on dust lifting.

In fact, studies by \citet{Renno2008, Holstein2010} showed that electrostatic forces can be beneficial to dust lifting by lowering the threshold physical force, that needs to be applied to a soil particle by an impactor, to break the cohesive chains binding it to the soil. \citet{Kok2008} used a model to demonstrate that saltating grains can build up charge during the saltation process due to repeated impacts into the soil and \citet{Schmidt1998, Zheng2003, Kruss2021} found, that these electrified particles can lead to an electric field building up between the saltators and the ground, benefitting further lifting. This finding is also supported by field study by \citet{Esposito2016}, who measured up to a ten fold increase in lifted dust when an electric field is applied.
We connect to these earlier studies by investigating the charges in detail that manifest on individual grains which are ejected in saltating impacts under laboratory conditions.

\section{Experiment}\label{sec:Exp}
We set up an experiment to simulate saltating impacts and observe the ejecta on a micro-scale. We start with a small number of sand-sized grains which impact a dust bed while strong horizontal electric field is applied. The motion of ejected particles is observed with a high speed camera. Within the homogeneous electric field, charges move on parabolic trajectories. With measured particle size and acceleration the particle charge is determined. 

\begin{figure}
	\includegraphics[width=\columnwidth]{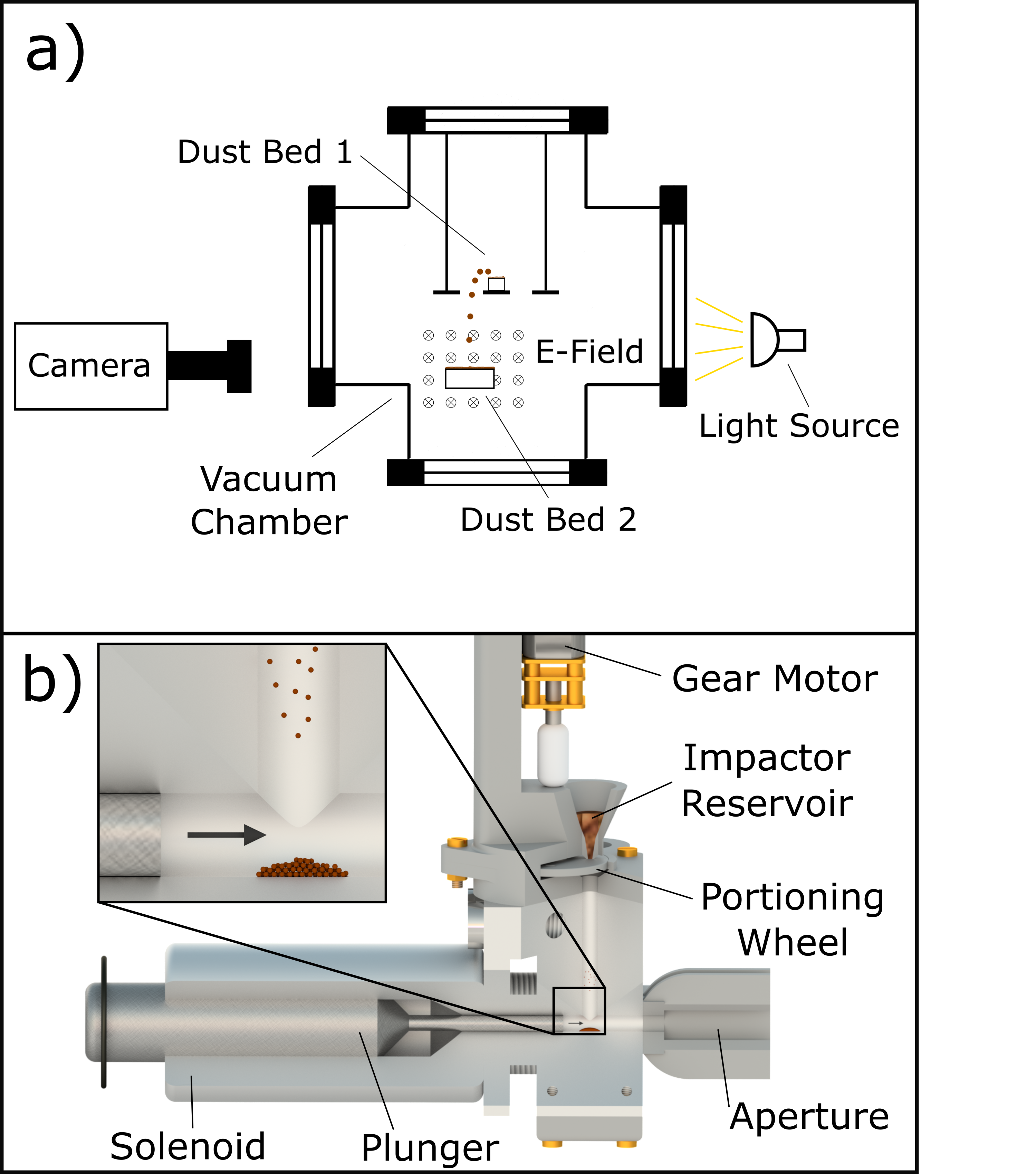}
	\caption{\label{fig.aufbau}Experiment Setup, where \textbf{a)} shows the vacuum chamber with the first and second dust     bed, electric field and positioning of the light source and \textbf{b)} shows the launching mechanism for the primary      impactors in detail.}
\end{figure}

\begin{figure}
	\includegraphics[width=\columnwidth]{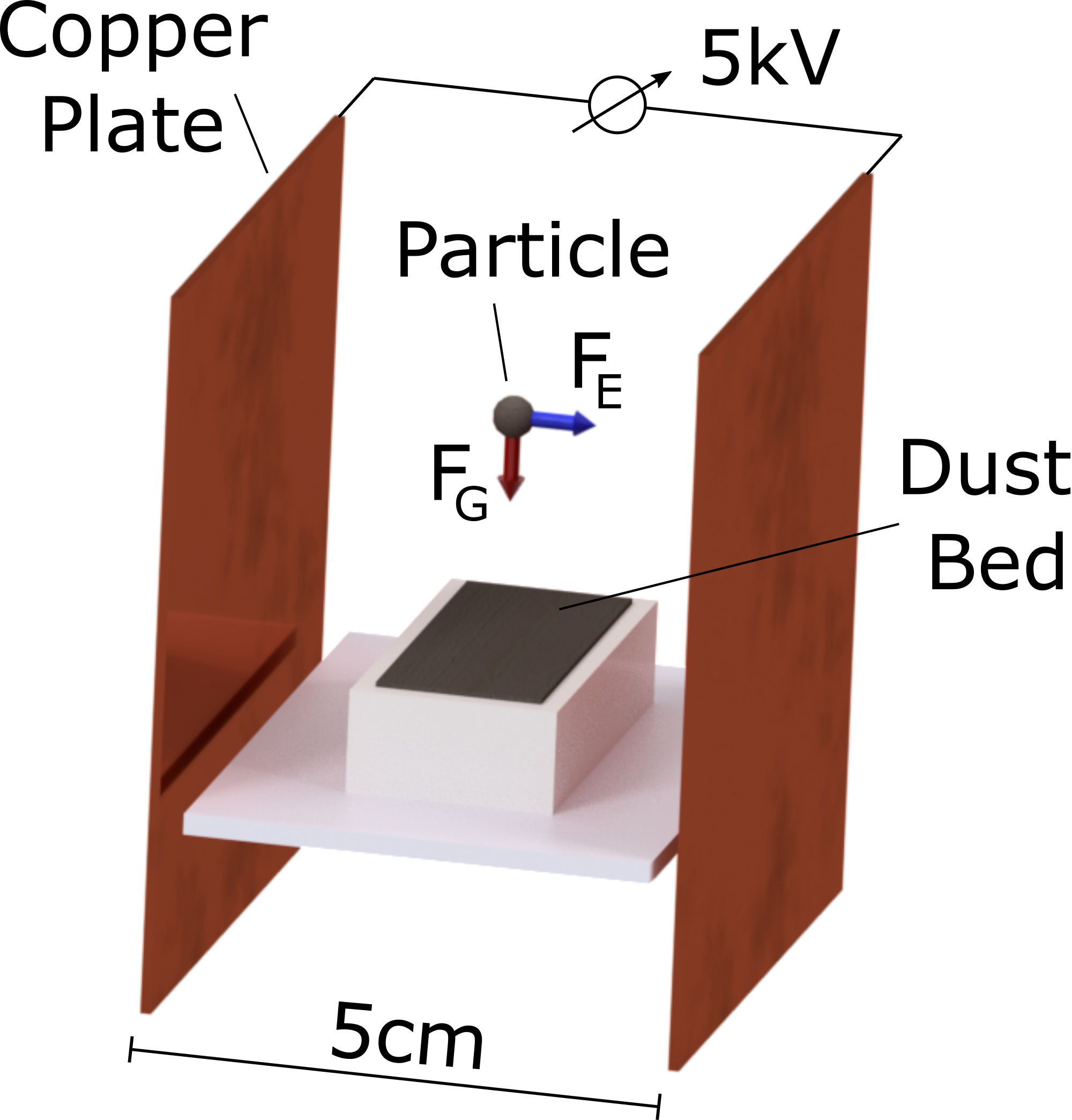}
	\caption{\label{fig.aufbau2}Sketch of the setup with the capacitor consisting of two copper plates to the left and right of the dustbed. The capacitor is operated at a voltage of 5kV and a distance of 5cm. The forces acting upon a (charged) particle are shown with a blue (electrostatic Force) and red arrow (gravity) respectively on an exemplary particle.}
\end{figure}

\subsection{Impact setup}
The basic setup is a slightly modified version of the one described in \citet{Becker2023}. The  experiments are carried out in a vacuum chamber at a pressure of $10^{-3}$ mbar. The pressure was chosen low in order to minimize the effect of gas drag when analyzing particle acceleration.
For the experiment itself, sand grains of sizes from 180 to 250 \textmu{}m have been used as primary saltators. They were launched at velocities of 1.04m/s$\pm$ 0.2m/s, using the accelerator shown in fig. \ref{fig.aufbau}\textbf{b)}. In each run, 20-50 saltators impact the top dust bed ("Dust Bed 1" in fig. \ref{fig.aufbau}\textbf{a)}). When impacting the dust bed, dust and sand is ejected, of which some part is falling down to the side of the bed. At a vertical distance of about 5 cm below the first dust bed another particle bed is placed ("Dust Bed 2" in fig. \ref{fig.aufbau}\textbf{a)}), so that the secondary particles that are ejected to the side in the first impact can re-impact in the second dust bed, yet again with velocities of $\approx$1 m/s. The lower bed is placed within a plate capacitor with a field strength of 100 kV/m, so that charged particles are accelerated in horizontal direction depending on the polarity of the charge they carry. The setup is shown in fig. \ref{fig.aufbau2}. Trajectories of particles ejected within the field are captured at 2000 fps with a NAC MEMRECAM HX-3 high speed camera. 

We chose a setup with two dust beds on top of each other mostly for technical reasons. 
This is not so different from a natural saltation cascade though, where primary saltators impact the soil and then the ejected particles work as next saltators. The impact angle changes from oblique, or 18.8$^\circ \pm$2.5$^\circ$ to near 90$^\circ$. 

\subsection{Soil: Martian simulant}
In our experiment we use MGS (Martian Global Simulant) as soil simulant \citep{Cannon2019}. This was also used in our former works \citep{Becker2022, Becker2023}. Again, we used a bimodal distribution of clay ($<$20 \textmu{}m) and sand sized ($\approx$100 \textmu{}m) particles. We used this distribution, since during natural saltation events there is a continuous mixing of larger saltators and finer dust, sedimented from the atmosphere, as well. We use a volume ratio of 1:1 of small and large particles. This ratio is chosen somewhat arbitrary, but keep it here so that all our works on this topic are comparable.

\section{Results}

\begin{figure}
	\includegraphics[width=\columnwidth]{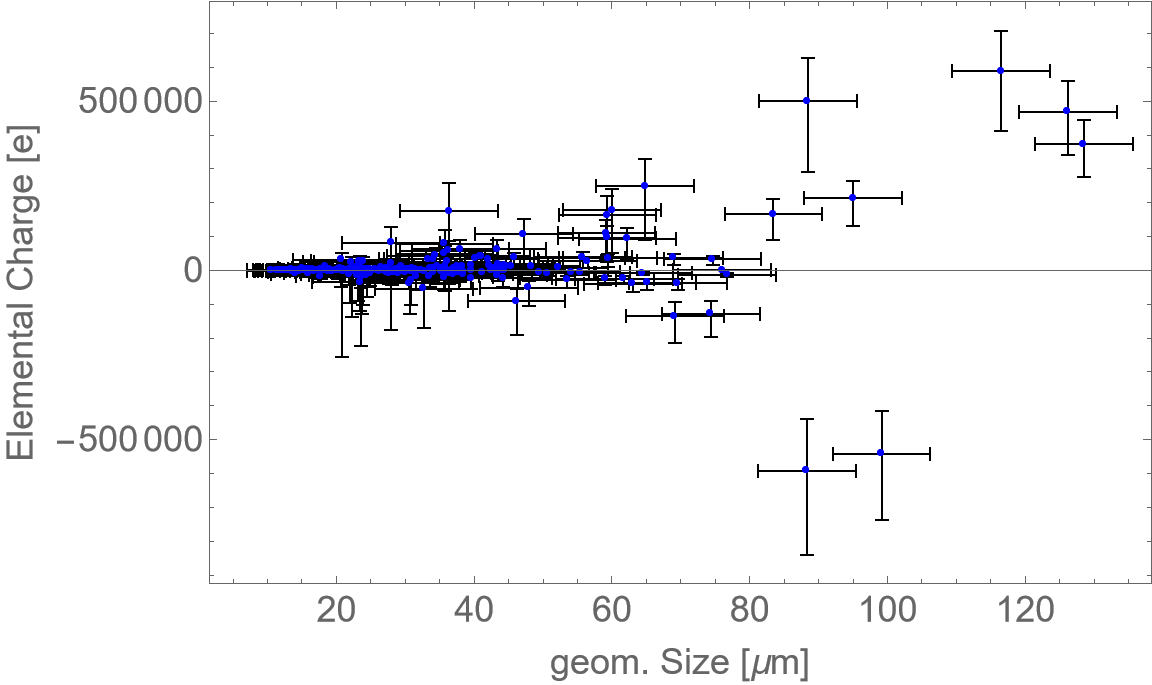}
	\includegraphics[width=\columnwidth]{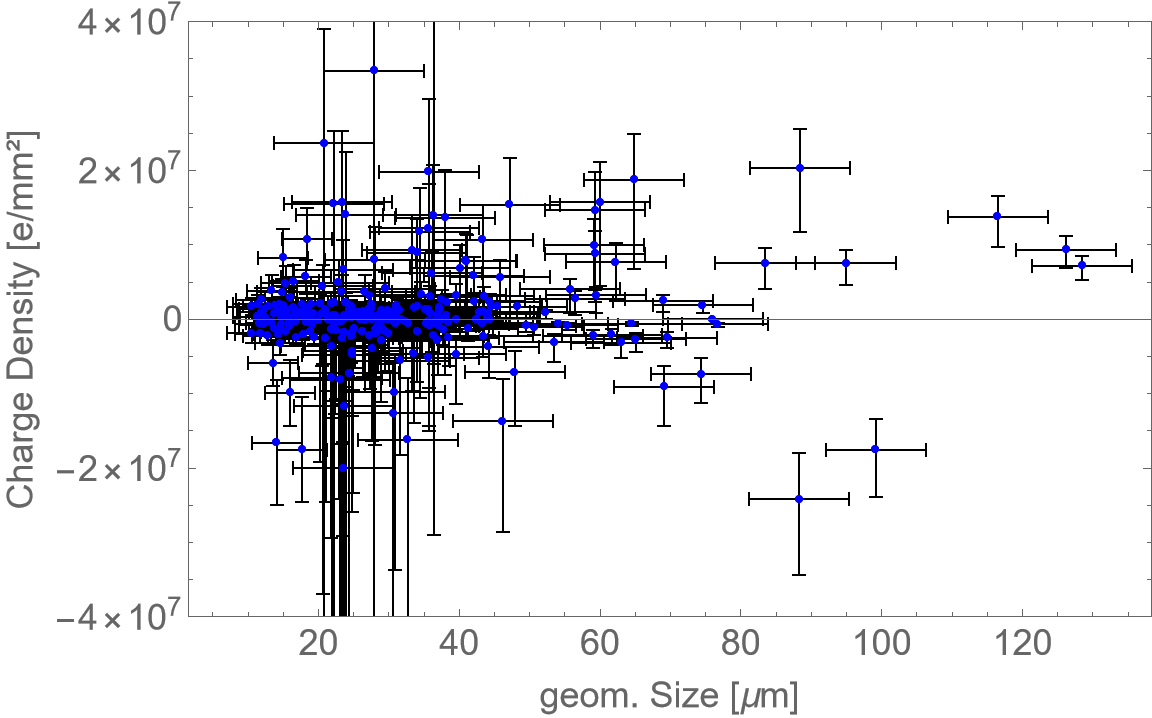}
        \includegraphics[width=\columnwidth]{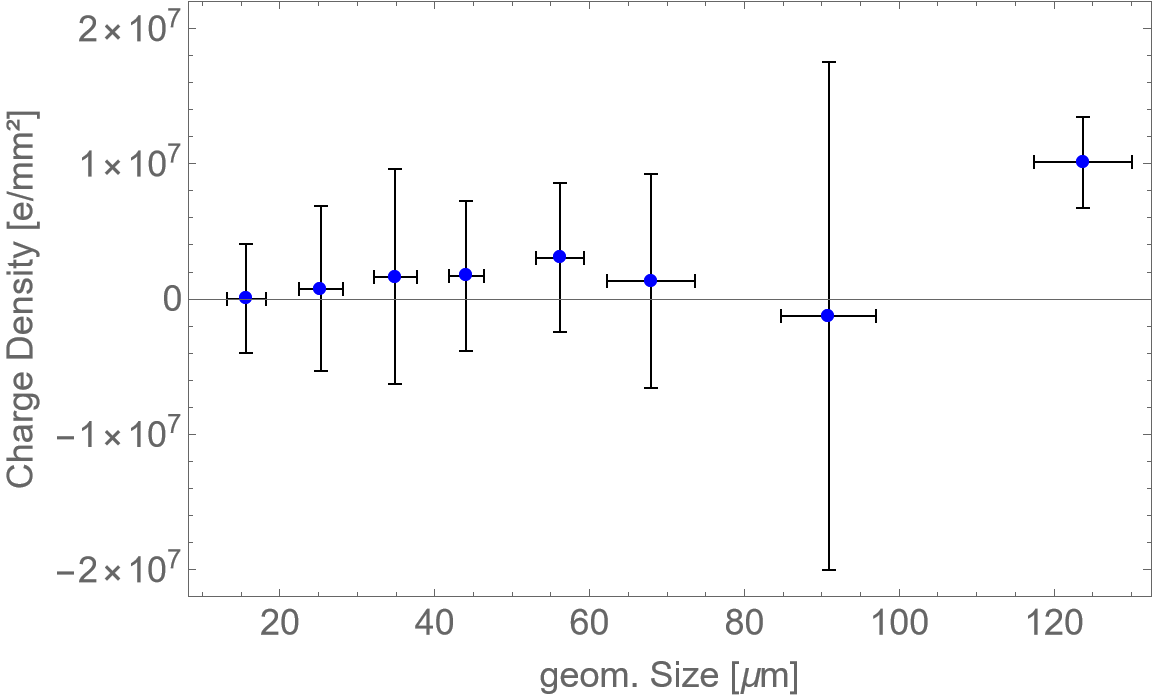}
	\caption{\label{evolution} Absolute charge (top) and charge density of ejeccted grains (middle). Bottom: binned data.}
\end{figure}


Figure \ref{evolution} shows the charges over particle size for all 283 individual particles that have been observed. The top plot shows the absolute values. The middle plot shows the charge densities related to the surface and the bottom plot shows the charge densities binned.

There is a small bias of the average charge toward positive values. As the charge density shows little size dependence, we give an average of all net particle charges of $1.05\cdot 10^{6}$ e/mm$^2$ with an error of the average of $2.3\cdot 10^{4}$ e/mm$^2$. The positive bias is about 16\% of the standard deviation of $6.5\cdot 10^{6}$ e/mm$^2$.
 
There are two potential reasons for the positive polarity bias. On one side, the bias might originate within the dust beds. E.g., net charge might already be present on the grains. However, as the samples are regularly prepared in ambient air, most of the charge should be equilibrated by conduction. The same argument of charge equilibration would apply for a net charge bias of the dust beds in general.

On the other side though, we have to note that we do not observe the grains smaller than a few micrometer, which were the focus of our first study \citep{Becker2022} and according to that, those particles are also liberated upon impact. Unfortunately, they are well beyond the resolution of our optics. Those small grains might account for the negative charge that is missing. Such reasoning would be in line with the usual findings that small grains in contact with larger grains collect negative and the larger ones positive charge. As we do not have many collisions between small and large grains, though, this usual size dependent polarity might not necessarily apply at the stage of particle ejection. 
So, at this point we can only speculate if the bias of charge polarity for the observed (large) particles is a consequence of the particle´s  ejection.

What we do not see is a significant size dependent charge polarity in the observed size range of about 10 microns to 100 microns. However, we definitely do see large net charge densities on ejected grains.

\section{Conclusion}

We analyzed the charge of individual particles ejected in simulated sand grain impacts under laboratory conditions. 
Particle size distribution and composition are chosen to simulate Martian soil, though the findings might also apply to other rocky planetary surfaces.
Ejected particles can acquire large net charges during ejection. We could only observe particles larger than 10 microns. At these sizes, there is a small bias toward positive polarity. The negative charges might be tied to smaller grains below the resolution threshold. The charge variations are large though and not limited to positive charges. 

If this interpretation holds and the polarity bias comes from the grain size, then ejection of grains would readily lead to a vertical electric field as the small grains become entrained into the atmosphere. Then, no collisional evolution would be necessary to induce a size dependent polarity bias to generate electric fields. Such a collisional evolution among freely colliding particles might be difficult anyway, as the dust grains usually tend to stick to larger grains or among themselves.

In any case, grains regularly have large net charges already right after their ejection. 
If electric fields are present or evolve during a saltation cascade by size dependent collisions, then a significant amount of grains are subject to lifting or restraining forces, which might shift thresholds for further saltation.



\section{Acknowledgments}
This project has received funding from the European Union’s Horizon 2020 research and innovation program under grant agreement No 101004052.
F.C. Onyeagusi is supported by DLR Space Administration with funds provided by the Federal Ministry for Economic Affairs and Climate Action (BMWK) under grant number DLR 50 WM 2142.

\bibliography{bibbi}
\bibliographystyle{aasjournal}

\end{document}